\documentclass[letterpaper,journal]{IEEEtran}
\usepackage{iftex}
\ifXeTeX
  \defaultfontfeatures{Ligatures=TeX}
  \makeatletter
  \renewcommand{\IEEEPARstart}[2]{#1#2}
  \makeatother
\fi
\usepackage{amsmath,amsfonts}
\usepackage{algorithmic}
\usepackage{algorithm}
\usepackage{array}
\usepackage[caption=false,font=normalsize,labelfont=sf,textfont=sf]{subfig}
\usepackage{textcomp}
\usepackage{stfloats}
\usepackage{url}
\usepackage{graphicx}
\usepackage{xcolor}
\usepackage{bm}
\usepackage{upgreek}
\usepackage{booktabs}
\usepackage{cite}
\usepackage{balance}
\usepackage{capt-of}
\newcommand{\resultfigheight}{0.13\textheight}

\begin{document}

\title{Neural Network-Enabled Codebook Design for Phased Array Calibration with Arbitrary Array Sizes}

\author{Xiangyu Chen, Hao Sun, Weiming Wang, Yuanan Liu, Wei Fan%
\thanks{Xiangyu Chen, Weiming Wang, and Yuanan Liu are with Beijing University of Posts and Telecommunications, Beijing 100876, China (e-mail: 2019212267@bupt.edu.cn; wangwm@bupt.edu.cn; yuliu@bupt.edu.cn).}%
\thanks{Wei Fan is with the National Mobile Communications Research Laboratory, School of Information Science and Engineering, Southeast University, Nanjing 210096, China, and also with the Purple Mountain Laboratories, Nanjing 211111, China (e-mail: weifan@seu.edu.cn).}%
\thanks{Hao Sun is with the China Academy of Information and Communications Technology, Beijing, China (e-mail: sunhao@caict.ac.cn).}}

\markboth{Journal of \LaTeX\ Class Files,~Vol.~14, No.~8, August~2021}%
{Shell \MakeLowercase{\textit{et al.}}: A Sample Article Using IEEEtran.cls for IEEE Journals}

\maketitle

\begin{abstract}

Array calibration is critical to achieving accurate beamforming in millimeter-wave (mmWave) antenna-in-package (AiP) phased arrays, where over-the-air (OTA) calibration in {\footnotesize ALL--ON} mode is a standard requirement. For practical calibration measurements, two core metrics are paramount: efficiency (defined by measurement time) and reliability (robustness, governed by the condition number of the phased array calibration codebook). In this work, we propose a neural network-enabled codebook generation method for phased array calibration compatible with arrays of arbitrary sizes. Codebooks generated via the proposed method achieve low condition numbers while requiring the minimum number of measurements, outperforming state-of-the-art calibration approaches. Practical measurements on a 26-GHz AiP phased array validate the effectiveness and robustness of the proposed method, with superior performance in both array calibration accuracy and beamforming quality.

\end{abstract}

\vspace{-5pt}
\section{Introduction}

\IEEEPARstart{M}{illimeter} wave (mmWave) antenna in packages (AiPs) have been widely adopted in mmWave communication and radar systems~\cite{7967694}. However, without proper calibration, AiP arrays suffer from gain degradation and pointing errors, which significantly deteriorate beamforming performance. Because AiP arrays usually do not provide independent RF ports for direct cable access, phased array calibration is typically performed in an anechoic chamber through an over-the-air (OTA) approach~\cite{he2019fast}. Therefore, efficient and practical OTA calibration is critical for ensuring the performance of mmWave AiP phased arrays.

In OTA array calibration, {\footnotesize ON--OFF} measurement schemes were widely adopted, where only a single antenna element is activated for measurement while the remaining elements are turned off. However, it was demonstrated that {\footnotesize ALL--ON} calibration, in which all array elements remain active and phase toggling is applied during measurement, achieves superior calibration performance for mmWave AiP arrays because it more closely matches the actual operating mode of phased arrays~\cite{9335235,9423585,10598309}. \par

Existing {\footnotesize ALL--ON} calibration methods can be divided into power-only and complex-signal-based methods. Power-only methods, such as the rotating element electric field vector (REV) method~\cite{mano1982method}, fast amplitude-only methods~\cite{7745942}, and related variants~\cite{9445668,8051943}, use simple power measurements but may suffer from low efficiency and accuracy for large-scale arrays~\cite{10326459}. For complex-signal-based methods, the key metrics are the excitation-matrix condition number and the number of measurements. For an $N$-element AiP phased array, $M$ codebooks and the corresponding $M$ measurements are theoretically sufficient for calibration through linear inversion, where $M\ge N$ is required. Representative methods include the Hadamard-based method~\cite{552217}, the improved Hadamard method~\cite{8693562}, the recursive-product-based method~\cite{7902129}, and the Fourier-structured excitation matrix method~\cite{10015637}. However, these methods are generally restricted by array size, matrix structure, or operating conditions. Moreover, recent beam-steering~\cite{9423585} and near-field methods~\cite{tang2023nearfield_extrapolation} still do not directly provide finite-bit low-condition-number codebooks for arbitrary array sizes. Artificial intelligence (AI)-based methods have also been investigated for phased-array calibration~\cite{9931606,9760077,9977899,10919091,9260769}, but some of them are difficult to scale to large arrays, whereas others still lack sufficiently comprehensive measurement validation.

Therefore, a practical {\footnotesize ALL--ON} calibration method that simultaneously provides high measurement efficiency with the minimum number of measurements, i.e., $M=N$, a low excitation-matrix condition number close to $1$, and compatibility with arbitrary array sizes is still needed. To address this issue, this letter proposes a neural-network-enabled excitation-matrix generation method for phased-array calibration. The proposed method achieves near-minimum condition numbers using the minimum number of measurements ($M=N$) and is applicable to arrays of arbitrary sizes.
\vspace{-5pt}
\vspace{-5pt}
\section{Codebook Analysis and Network Design}
\vspace{-5pt}
\subsection{Problem Statement}

In {\footnotesize ALL--ON} OTA calibration, multiple array elements are excited simultaneously according to a predefined calibration codebook. Let $\mathbf{A}\in\mathbb{C}^{M\times N}$ denote the complex-valued calibration codebook matrix, whose rows correspond to measurement states and whose columns correspond to active array elements. Here, $N$ is the number of active array elements and $M$ is the number of measurements. Let $\mathbf{x}\in\mathbb{C}^{N\times 1}$ denote the unknown complex calibration coefficients of the active elements, $\mathbf{y}\in\mathbb{C}^{M\times 1}$ denote the measured observation vector, and $\mathbf{n}\in\mathbb{C}^{M\times 1}$ denote the measurement-noise vector. The calibration model is
\begin{equation}
\label{eq:cal_model}
\mathbf{y}=\mathbf{A}\mathbf{x}+\mathbf{n},
\end{equation}
To emphasize the minimum-measurement setting, we first consider $M=N$. If $\mathbf{A}$ is nonsingular, the estimated calibration vector $\hat{\mathbf{x}}$ is recovered by
\begin{equation}
\label{eq:cal_inverse}
\hat{\mathbf{x}}=\mathbf{A}^{-1}\mathbf{y},
\end{equation}
where $\mathbf{A}^{-1}$ is the inverse of $\mathbf{A}$. The reconstruction error is then
\begin{equation}
\label{eq:cal_error}
\hat{\mathbf{x}}-\mathbf{x}=\mathbf{A}^{-1}\mathbf{n},
\end{equation}
which gives
\begin{equation}
\label{eq:error_bound}
\|\hat{\mathbf{x}}-\mathbf{x}\|\leq \|\mathbf{A}^{-1}\|\,\|\mathbf{n}\|.
\end{equation}
Here, $\|\cdot\|$ denotes the Euclidean norm for vectors and the induced $2$-norm for matrices. Equation~(\ref{eq:error_bound}) shows that measurement noise can be amplified by $\mathbf{A}^{-1}$. The condition number of $\mathbf{A}$ is
\begin{equation}
\label{eq:kappa_def}
\kappa(\mathbf{A})=\|\mathbf{A}\|\,\|\mathbf{A}^{-1}\|.
\end{equation}
For a small observation perturbation $\delta\mathbf{y}$ and the corresponding coefficient perturbation $\delta\mathbf{x}$, the standard normwise perturbation analysis of linear systems gives~\cite{higham1995condition}

\begin{equation}
\label{eq:relative_error}
\frac{\|\delta\mathbf{x}\|}{\|\mathbf{x}\|}\leq \kappa(\mathbf{A})\frac{\|\delta\mathbf{y}\|}{\|\mathbf{y}\|}.
\end{equation}
Thus, a smaller $\kappa(\mathbf{A})$ implies more stable inversion and higher calibration accuracy. Accordingly, codebook design should minimize $\kappa(\mathbf{A})$ under practical hardware constraints. For $M<N$, the system is underdetermined, so $\mathbf{A}^{-1}$ does not exist and unique calibration recovery fails.

It was shown in~\cite{7902129} that matrix conditioning depends strongly on matrix structure. In this recursive-product-based method, a low-condition-number $N\times N$ codebook, corresponding to the minimum-measurement case $M=N$, can be directly constructed only when $N$ is factorized into the orders of the basic matrices $\mathbf{A}_2$, $\mathbf{A}_3$, $\mathbf{A}_5$, and Hadamard matrices. Otherwise, the problem must be embedded into a larger matrix, which increases the number of measurements to $M>N$; for example, for an $86$-element array, four dummy elements are introduced, increasing $M$ from $86$ to $90$. Therefore, practical {\footnotesize ALL--ON} OTA calibration requires low-condition-number codebooks while keeping $M$ as close to $N$ as possible.
\vspace{-5pt}
\vspace{-5pt}
\subsection{Proposed Algorithm}

\begin{figure}[!t]
\centering
\includegraphics[width=\columnwidth,height=0.32\textheight,keepaspectratio]{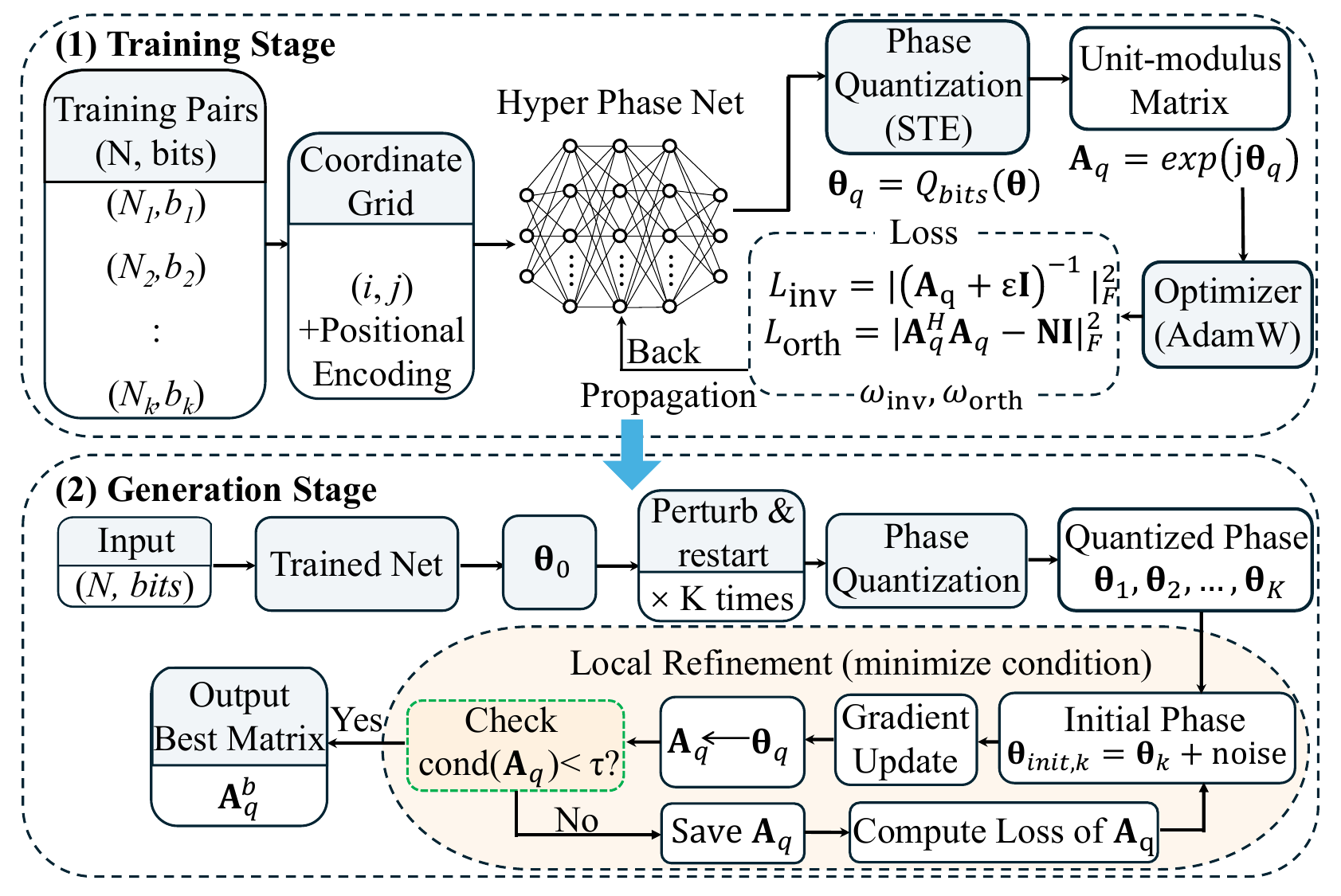}
\vspace{-20pt}
\caption{Proposed framework for calibration codebook generation.}
\label{fig:net_model}
\end{figure}

\begin{figure}[!t]
\centering
\includegraphics[width=1\columnwidth,height=0.5\textheight,keepaspectratio]{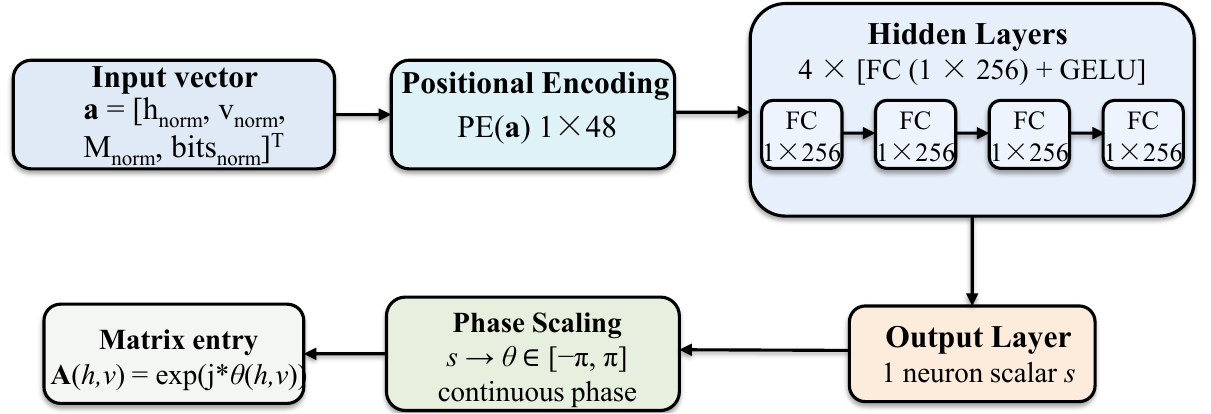}
\vspace{-20pt}
\caption{Structure of the hyper phase network.}
\label{fig:mesh_model}
\end{figure}

As shown in Fig.~\ref{fig:net_model}, the proposed method consists of training and generation stages. During training, the network learns the phase of each codebook entry from its row-column coordinate and array configuration. During generation, the trained network outputs an initial phase matrix for a target $(N,\mathrm{bits})$ pair, where $\mathrm{bits}$ is the phase-quantization bit number, and then refines it under the finite-bit phase constraint.

For the minimum-measurement setting $M=N$, the \emph{Coordinate Grid} enumerates the $N^2$ entries of the square codebook matrix. Each entry is represented by
\begin{equation}
\mathbf{a}=\left[h_{\mathrm{norm}},v_{\mathrm{norm}},N_{\mathrm{norm}},\mathrm{bits}_{\mathrm{norm}}\right]^{\top},
\end{equation}
where $(\cdot)^{\top}$ denotes transpose. The terms $h_{\mathrm{norm}}$ and $v_{\mathrm{norm}}$ denote the normalized row and column indices of the codebook entry, respectively, while $N_{\mathrm{norm}}$ and $\mathrm{bits}_{\mathrm{norm}}$ denote the normalized array size and phase resolution. Let $a_m$ denote the $m$th component of $\mathbf{a}$. This representation allows the same network to be shared across different array sizes and phase resolutions.

$\mathrm{PE}(\mathbf{a})$ denotes the positional-encoding mapping of the input vector $\mathbf{a}$, which maps the input vector to Fourier features through the \emph{Positional Encoding} block in Fig.~\ref{fig:mesh_model}:
\begin{equation}
\mathrm{PE}(\mathbf{a})=
\left[
\sin(2^{\ell}\pi a_{m}),\,
\cos(2^{\ell}\pi a_{m})
\right]_{m=1,\ldots,4}^{\ell=0,\ldots,5},
\end{equation}
where $m$ and $\ell$ index the input component and frequency scale, respectively. The encoding gives a $1\times48$ feature vector, which is fed into a shared MLP with four 256-neuron hidden layers and GELU activation. The scalar output is scaled to $[-\pi,\pi]$ as $\theta(h,v)$, and evaluating all $N^2$ entries forms $\bm{\uptheta}\in\mathbb{R}^{N\times N}$.

During training, the continuous phase matrix $\bm{\uptheta}$ is quantized elementwise by the $\mathrm{bits}$-bit phase quantizer $Q_{\mathrm{bits}}(\cdot)$:
\begin{equation}
\bm{\uptheta}_{\mathrm{q}}=Q_{\mathrm{bits}}(\bm{\uptheta}),
\end{equation}
where the subscript $\mathrm{q}$ denotes quantization and each entry is mapped to the nearest realizable discrete phase state. Since quantization is non-differentiable, the straight-through estimator (STE) in Fig.~\ref{fig:net_model} is used for backpropagation. The quantized phase matrix is then converted into the unit-modulus calibration codebook
\begin{equation}
\mathbf{A}_{\mathrm{q}}=\exp(j\bm{\uptheta}_{\mathrm{q}}),
\end{equation}
where $j$ is the imaginary unit and the exponential is applied elementwise. The training loss on $\mathbf{A}_{\mathrm{q}}$ is
\begin{equation}
L_{\mathrm{inv}}=\left\|(\mathbf{A}_{\mathrm{q}}+\epsilon\mathbf{I})^{-1}\right\|_{F}^{2},
\end{equation}
\begin{equation}
L_{\mathrm{orth}}=\left\|\mathbf{A}_{\mathrm{q}}^{H}\mathbf{A}_{\mathrm{q}}-N\mathbf{I}\right\|_{F}^{2},
\end{equation}
\begin{equation} 
L=\omega_{\mathrm{inv}}L_{\mathrm{inv}}+\omega_{\mathrm{orth}}L_{\mathrm{orth}},
\end{equation}
where $\|\cdot\|_{F}$ denotes the Frobenius norm, $\epsilon$ is a small positive constant for numerical regularization, $\mathbf{I}$ is the identity matrix, and $(\cdot)^{H}$ denotes conjugate transpose. Here, $L$ is the total training loss, $L_{\mathrm{inv}}$ penalizes poor invertibility, and $L_{\mathrm{orth}}$ drives $\mathbf{A}_{\mathrm{q}}^{H}\mathbf{A}_{\mathrm{q}}$ toward $N\mathbf{I}$ to encourage a low condition number. The weights are set to $\omega_{\mathrm{inv}}=1.0$ and $\omega_{\mathrm{orth}}=0.05$. The network parameters are optimized by AdamW with an initial learning rate of $5\times10^{-4}$, weight decay of $10^{-6}$, and a cosine annealing schedule. The training pairs are sampled from $N\in\{2,\ldots,256\}$ and $\mathrm{bits}\in\{2,\ldots,6\}$. The network is trained for $10{,}000$ epochs with a batch size of $8$, while $\epsilon$ is linearly decreased from $10^{-2}$ to $10^{-5}$ and gradient clipping with a maximum norm of $1.0$ is applied.

In the generation stage, the trained network first outputs an initial continuous phase matrix $\bm{\uptheta}_{0}$ for a target $(N,\mathrm{bits})$ pair. To reduce sensitivity to local minima after quantization, the \emph{Perturb \& restart} block performs $K$ random restarts, generating candidate quantized phase matrices $\bm{\uptheta}_{1},\bm{\uptheta}_{2},\ldots,\bm{\uptheta}_{K}$. For the $k$th candidate, with $k=1,\ldots,K$, local refinement starts from
\begin{equation}
\bm{\uptheta}_{\mathrm{init},k}=\bm{\uptheta}_{k}+\mathrm{noise},
\end{equation}
where $\bm{\uptheta}_{\mathrm{init},k}$ is the initial phase matrix for the $k$th local refinement and $\mathrm{noise}$ denotes a small random perturbation. The phase variables are then iteratively updated, quantized, and converted to $\mathbf{A}_{\mathrm{q}}$ so as to reduce $\mathrm{cond}(\mathbf{A}_{\mathrm{q}})$, where $\mathrm{cond}(\cdot)$ denotes the matrix condition number. The refinement terminates when $\mathrm{cond}(\mathbf{A}_{\mathrm{q}})<\tau$, where $\tau$ is a preset condition-number threshold, or when the maximum number of iterations is reached. The candidate with the smallest condition number is selected as the final codebook $\mathbf{A}_{\mathrm{q}}^{\mathrm{b}}$. In the experiments, $K=6$ is used, and $\tau$ is set to $1.5$, $1.3$, and $1.10$ for 4-6 bit codebooks, respectively.

Training, inference are performed offline on an Intel Core i7-14700K CPU and NVIDIA GeForce RTX 4090D GPU. Net has about $2.1\times10^{5}$ parameters, requiring 0.84 MB in FP32. In deployment, a library of codebooks is generated on the server, exported to the AiP and calibration-system control computer. Since each codebook depends only on the array size and AiP phase-step resolution, no real-time neural-network inference is required, the control computer only stores the precomputed complex codebook matrix and has negligible storage and computation burden. The codebook generation time increases with array size, reaching 32.58 ms for 256 elements.

\begin{figure}[!t]
\centering
\includegraphics[width=\columnwidth,height=0.22\textheight,keepaspectratio]{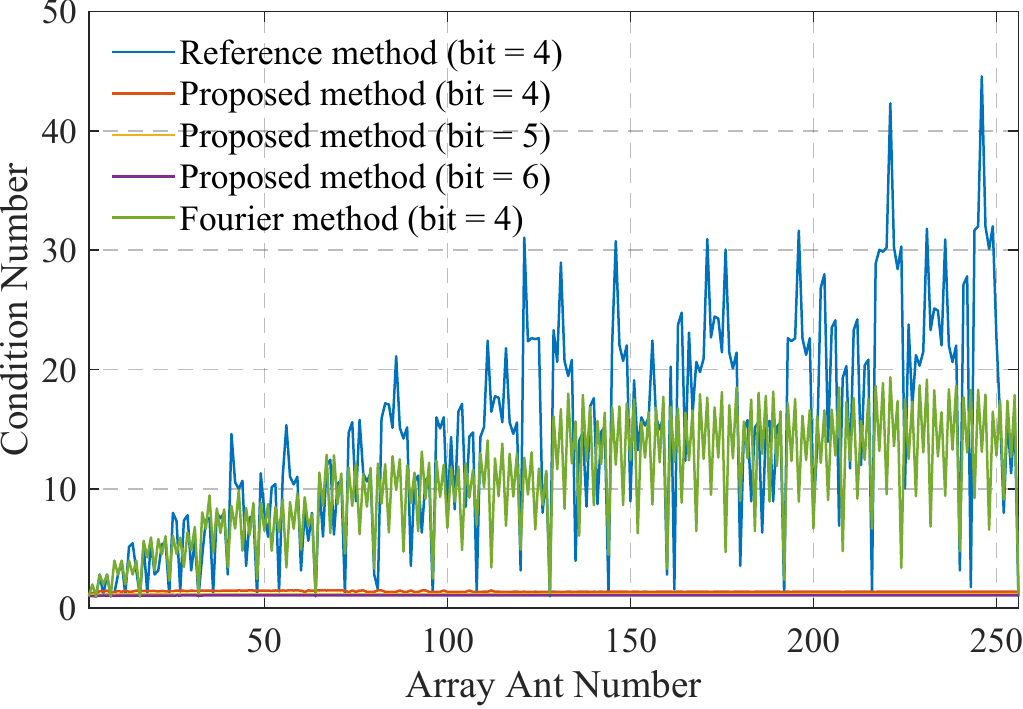}
\vspace{-12pt}
\caption{Condition numbers of the generated codebook matrices under different phase-quantization resolutions.}
\label{fig:condition_number}
\end{figure}
\vspace{-3pt}
\vspace{-5pt}
\vspace{-5pt}
\subsection{Numerical Simulations}

Fig.~\ref{fig:condition_number} compares the condition numbers of the proposed, recursive-product~\cite{7902129}, and Fourier-structured~\cite{10015637} codebooks for $M=N$. For $N=2$--$256$, the proposed method yields near-unit condition numbers for 4-, 5-, and 6-bit quantization, with average values of 1.41, 1.17, and 1.08 and maximum values of 1.45, 1.20, and 1.09, respectively, outperforming the reference method with limited measurements. Although the Fourier-structured method~\cite{10015637} can reach a near-unit condition number by increasing $M$, it must expand to the nearest power-of-two order. In contrast,~\cite{7902129} expands only to the nearest order composed of factors $2$, $3$, and $5$, requiring fewer extra measurements and keeping $M$ closer to $N$. Therefore,~\cite{7902129} is selected as the main reference, while~\cite{10015637} is included as an additional structured-codebook comparison. Moreover, the codebook performance depends only on the number of array elements and the phase-quantization resolution, rather than on the array layout and design.

\vspace{-10pt}
\begin{figure}[!h]
\centering
\includegraphics[width=\columnwidth,height=0.15\textheight,keepaspectratio]{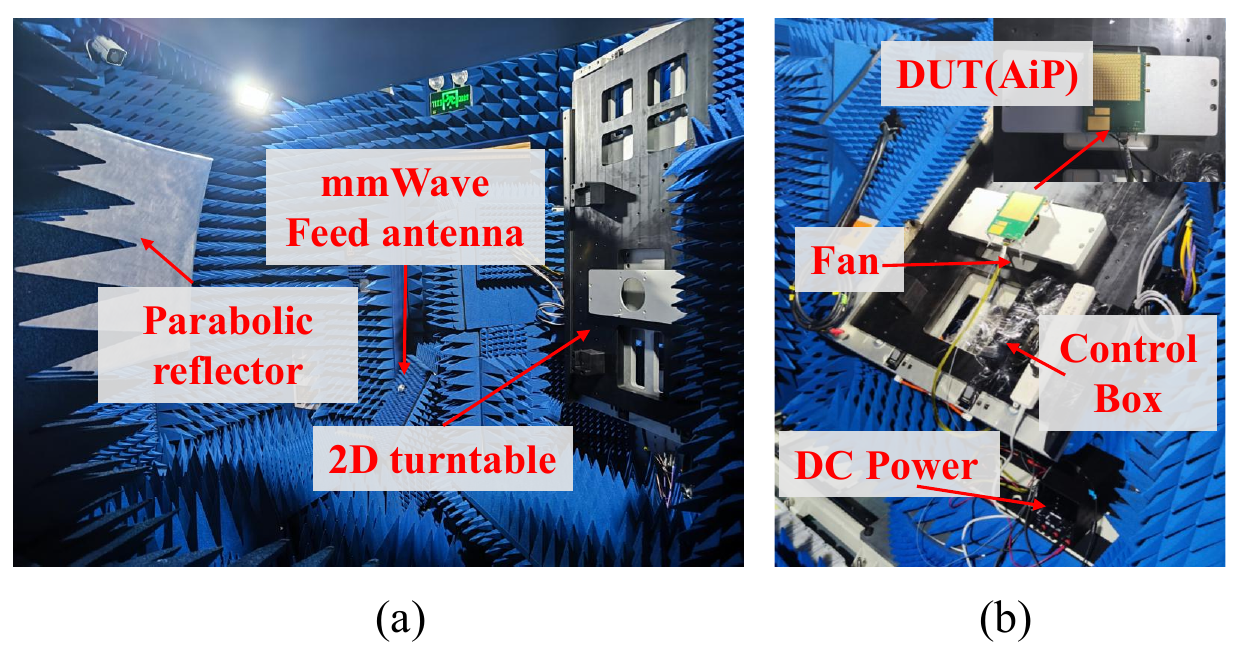}
\vspace{-12pt}
\caption{CATR environment and DUT mounting.}
\label{fig:real_mea}
\end{figure}
\vspace{-10pt}
\begin{figure}[!h]
\centering
\includegraphics[width=\columnwidth,height=0.15\textheight,keepaspectratio]{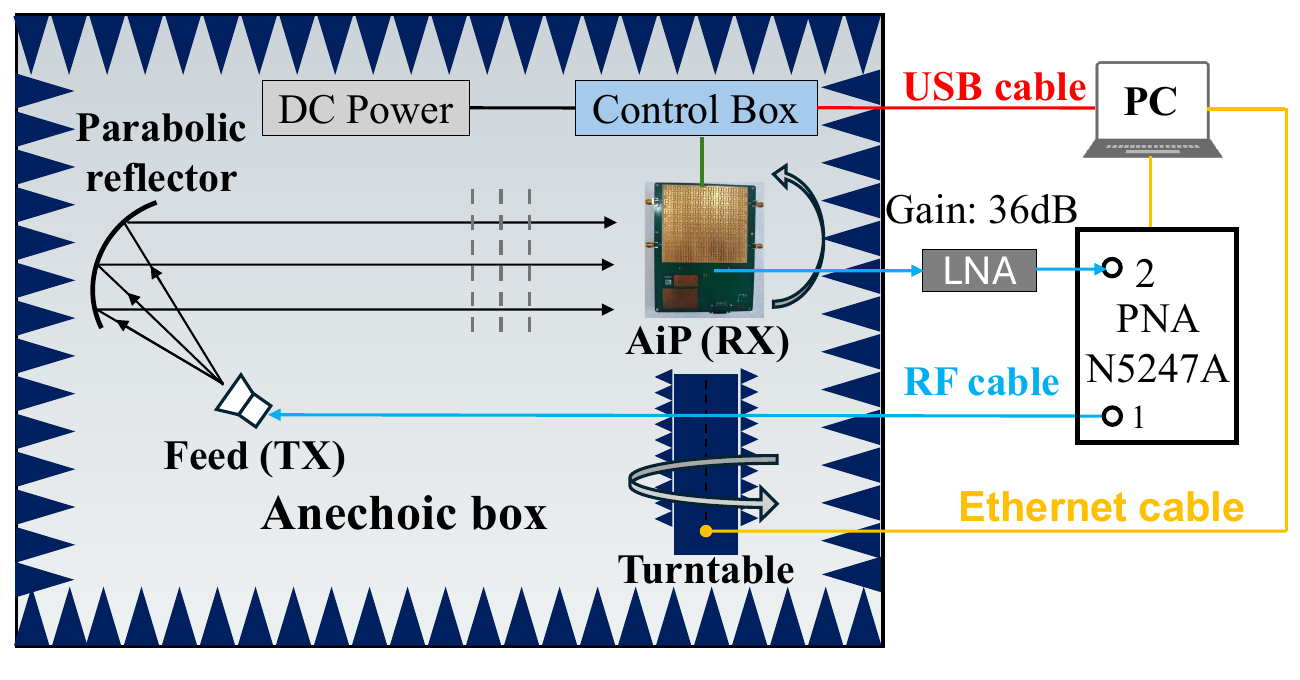}
\vspace{-12pt}
\caption{Complete OTA calibration setup.}
\label{fig:mea_model}
\end{figure}

\setlength{\textfloatsep}{4pt}
\begin{figure*}[!t]
\centering
\begin{minipage}[t]{0.49\textwidth}
\centering
\includegraphics[height=\resultfigheight,keepaspectratio]{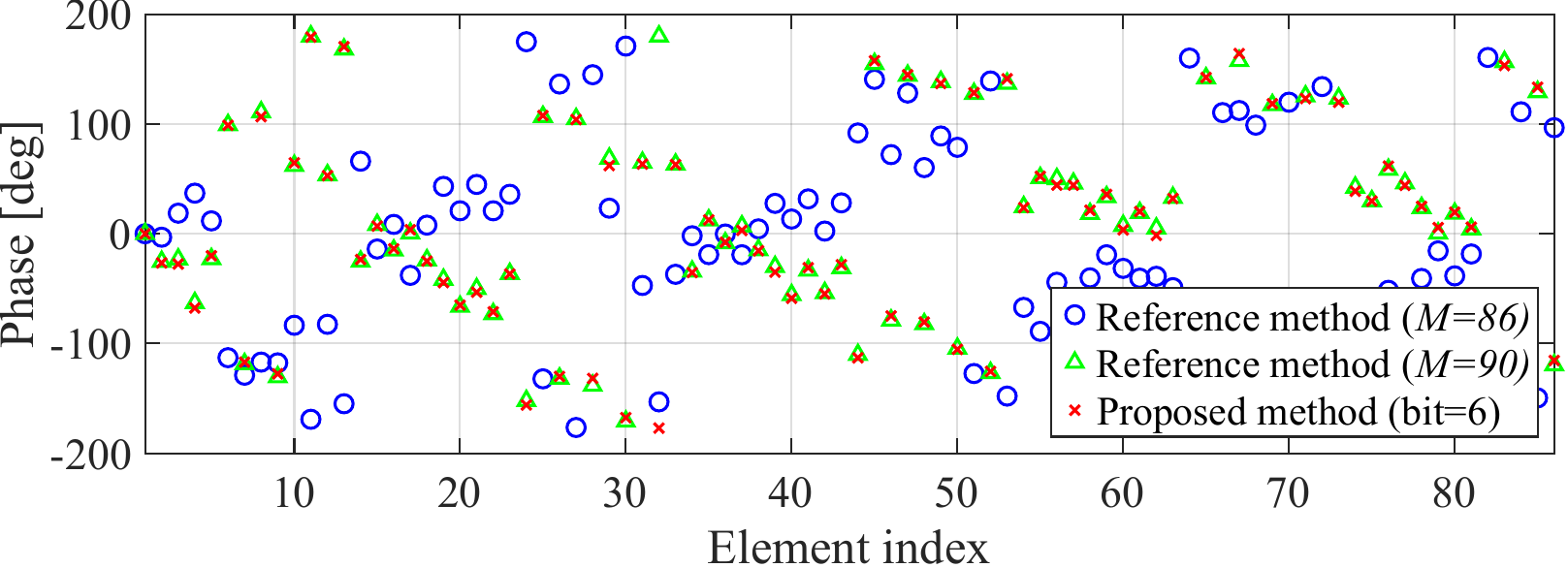}
\label{fig:phase}
\vspace{-16pt}

\centerline{\footnotesize (a)}
\vspace{0.25em}
\includegraphics[height=\resultfigheight,keepaspectratio]{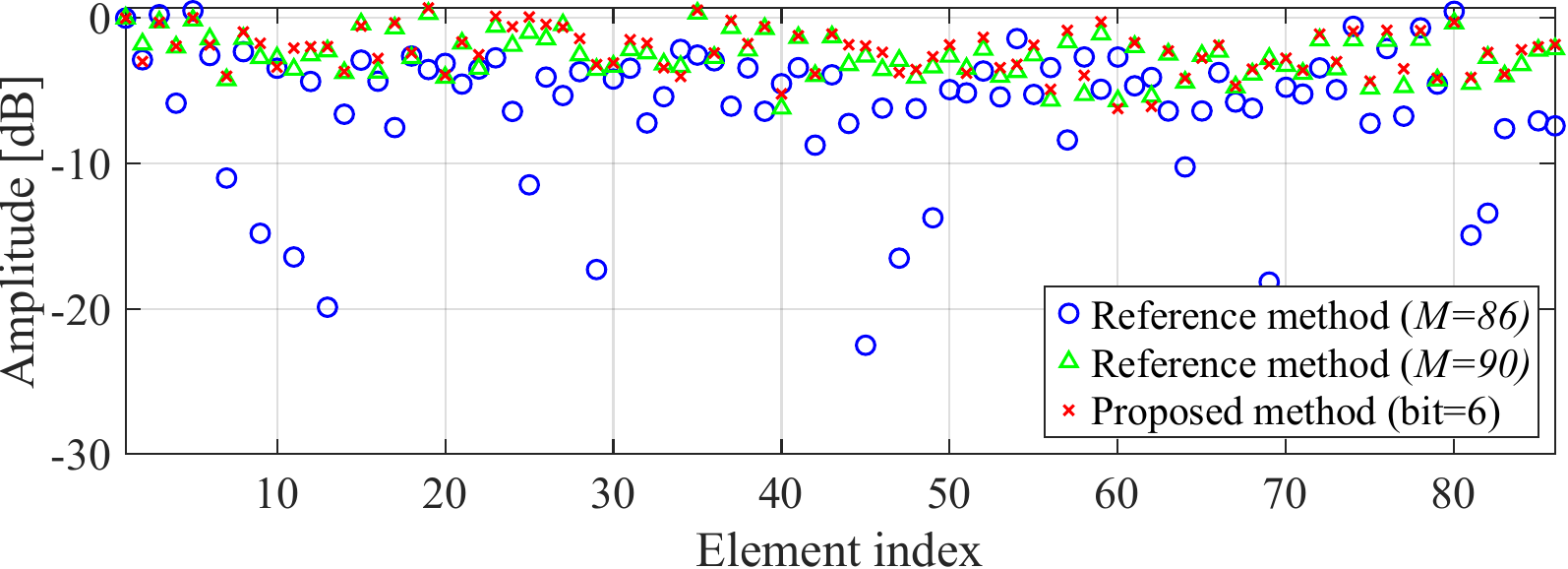}
\label{fig:amplitude}
\vspace{-16pt}
\centerline{\footnotesize (b)}
\vspace{-20pt}
\captionof{figure}{Calibration results: (a) phase and (b) amplitude.}
\label{fig:calibration_results}
\end{minipage}\hfill
\begin{minipage}[t]{0.49\textwidth}
\centering
\includegraphics[height=\resultfigheight,keepaspectratio]{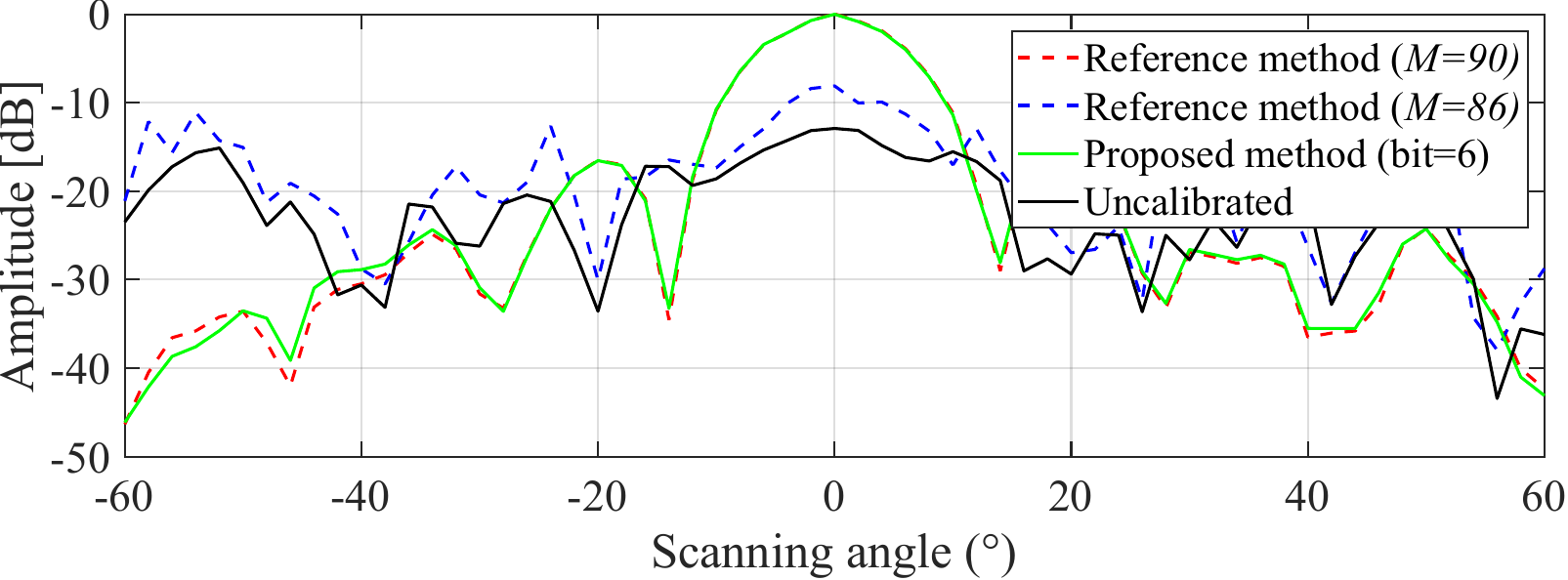}
\label{fig:Horizontal}
\vspace{-3pt}
\centerline{\footnotesize (a)}
\vspace{0.25em}
\includegraphics[height=\resultfigheight,keepaspectratio]{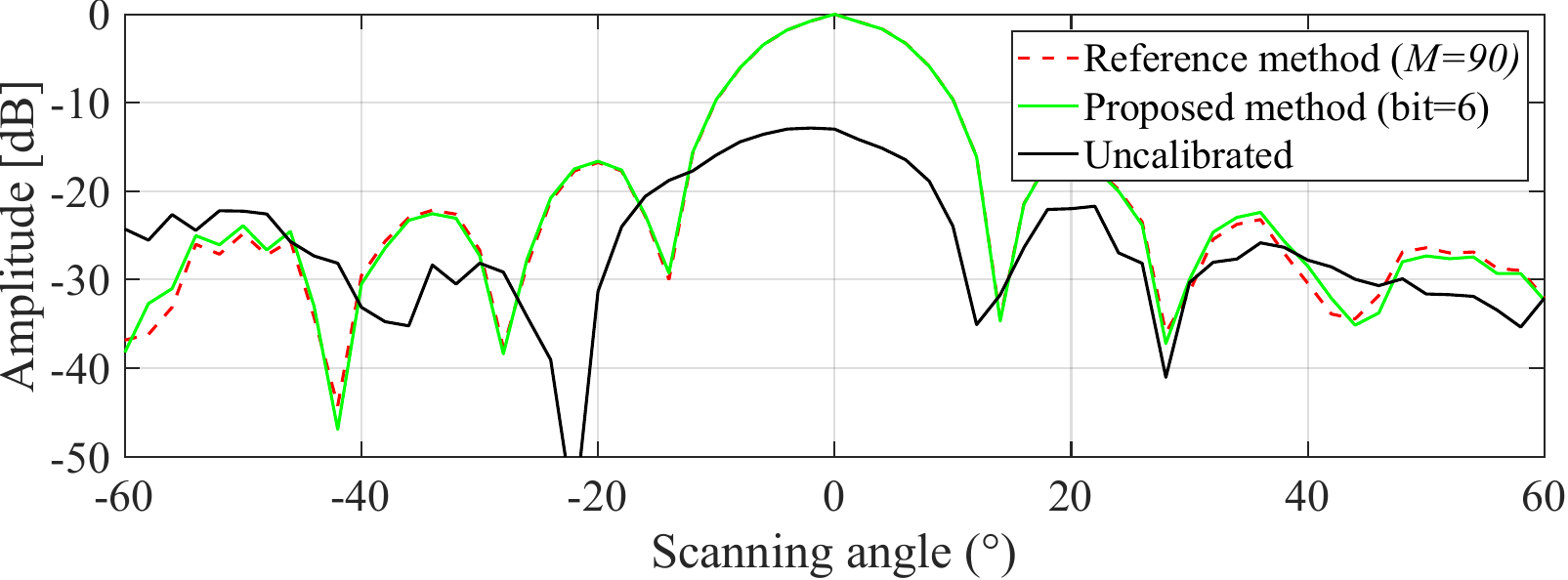}
\label{fig:Vertical}
\vspace{-3pt}
\centerline{\footnotesize (b)}
\vspace{-20pt}
\captionof{figure}{Beam-scanning results: (a) horizontal and (b) vertical planes.}
\label{fig:beamforming_results}
\end{minipage}
\end{figure*}
\vspace{-5pt}
\vspace{-5pt}
\section{Measurement Validation}

\subsection{Measurement Setup}

To verify that the proposed algorithm, experiments were conducted on a $26$-GHz mmWave AiP array in a compact antenna test range (CATR). As shown in Fig.~\ref{fig:real_mea}(a), a feed antenna and a parabolic reflector were used to generate plane-wave illumination toward the device under test (DUT), which was mounted on a two-dimensional turntable for both calibration and beam-scanning measurements. At $26$ GHz, the quiet-zone size of the CATR was $75\times 75$ cm, with amplitude and phase ripples of only $\pm 0.4$ dB and $\pm 4^{\circ}$.

As shown in Fig.~\ref{fig:real_mea}(b), the DUT was a $16\times 16$ AiP array with a radiating aperture of $9.8\times 8.9$ cm, which fits entirely within the quiet zone. Each element supports $6$-bit phase control over $0$--$360^{\circ}$ and amplitude control from $-31.5$ dB to $0$ dB with a step size of $0.5$ dB. Fig.~\ref{fig:mea_model} shows the complete OTA calibration setup. A Keysight PNA N5247A vector network analyzer (VNA) was used to acquire the complex transmission coefficient $S_{21}$ for each codebook state, with Port $1$ connected to the feed antenna and Port $2$ connected to the AiP through a low-noise amplifier (LNA) with a gain of $36$ dB. The center frequency, source power, and intermediate-frequency bandwidth of the VNA were set to $26$ GHz, $10$ dBm, and $10$ Hz, respectively. All equipment control is implemented in MATLAB, including the turntable, AiP and VNA.

For calibration, we selected the central $86$ elements as the active subarray to reduce measurement time, while the other elements were turned off and one defective element was excluded. This configuration keeps the calibrated aperture within the quiet zone and provides a challenging case because the reference codebook with $M=N=86$ has a relatively large condition number. Table~\ref{tab:method_comparison} lists the tested codebooks and also includes the condition-number comparison for the ideal array size $N=256$. For $N=256$, the reference method is equivalent to a standard Hadamard matrix and achieves a condition number of $1$, while the proposed codebook still maintains a near-unit condition number with the minimum number of measurements, with $M=N$. Moreover, the proposed method can maintain favorable performance for arbitrary numbers of antenna elements.
\vspace{-5pt}
\vspace{-5pt}
\begin{table}[htbp]
\centering
\caption{Condition Numbers for Calibration Methods at $N=$ $86$ and $256$}
\label{tab:method_comparison}
\footnotesize
\setlength{\tabcolsep}{4pt}
\begin{tabular}{@{}clcc@{}}
\toprule
$N$ & Method & $M$ & Condition number \\
\midrule
$86$ & Reference method~\cite{7902129}  & $86$ & $21.12$ \\
$86$ & Reference method & $90$ & $3.52$ \\
$86$ & Proposed method ($bit=6$) & $86$ & $1.08$ \\
$256$ & Reference method & $256$ & $1.00$ \\
$256$ & Proposed method ($bit=6$) & $256$ & $1.18$ \\
\bottomrule
\end{tabular}
\end{table}

\vspace{-5pt}
% \vspace{-5pt}
\subsection{Measurement Results}

Fig.~\ref{fig:calibration_results} compares the recovered phase and amplitude calibration coefficients for the reference methods and the proposed method with $86$ active elements. The reference method ($M=86$) shows low normalized amplitudes because its large condition number amplifies calibration errors, rather than indicating weak element radiation; its phase calibration also exhibits large errors. By contrast, the proposed method and the reference method ($M=90$) provide very close phase and amplitude calibration results, with mean absolute phase and amplitude differences of $2.42^\circ$ and $0.42$ dB, respectively.

Fig.~\ref{fig:beamforming_results} presents normalized beam-scanning results. In Fig.~\ref{fig:beamforming_results}(a), the reference method with $M=86$ shows severe degradation, confirming unreliable beamforming from the ill-conditioned codebook. The proposed codebook with $M=86$ closely follows the reference method with $M=90$. In Fig.~\ref{fig:beamforming_results}(b), these two effective calibrations also produce similar vertical-plane patterns, demonstrating comparable beam quality with fewer measurements.

These results indicate that the proposed low-condition-number codebook can achieve calibration and beam-scanning performance comparable to that of the reference method with $M=90$, while requiring only $M=N$ measurements. The CATR results therefore verify that accurate OTA calibration can be maintained with a reduced measurement burden by using the proposed codebook.
\vspace{-15pt}
% \vspace{-5pt}
\section{Conclusion}
In this letter, a neural-network-based codebook generation method has been proposed for OTA calibration of phased arrays with arbitrary array sizes. The generated constant-modulus codebooks exhibit low condition numbers, enabling accurate recovery of the initial element amplitudes and phases with only $M=N$ measurements. Measurements on a $26$-GHz AiP array show that the proposed method achieves calibration and beam-scanning performance comparable to that of a reference method requiring more measurements. This demonstrates the potential of the proposed method for efficient OTA calibration of mmWave phased arrays.

\balance
\bibliography{reference}

\end{document}